\documentclass{article}

\usepackage{microtype}
\usepackage{graphicx}
\usepackage{booktabs} % for professional tables
\usepackage{kotex}
\usepackage{multirow}

\usepackage{tabularx}
\usepackage{subcaption}
\usepackage{color}
\usepackage{soul}

\usepackage{hyperref}

\usepackage[accepted]{icml2018}

\icmltitlerunning{A Hybrid of Deep Audio Feature and i-vector for Artist Recognition} 

\begin{document}
\twocolumn[
\icmltitle{A Hybrid of Deep Audio Feature and i-vector for Artist Recognition}




\icmlsetsymbol{equal}{*}

\begin{icmlauthorlist}
\icmlauthor{Jiyoung Park}{nv}
\icmlauthor{Donghyun Kim}{nv}
\icmlauthor{Jongpil Lee}{ka}
\icmlauthor{Sangeun Kum}{ka}
\icmlauthor{Juhan Nam}{ka}
\end{icmlauthorlist}

\icmlaffiliation{nv}{NAVER Corp., Korea}
\icmlaffiliation{ka}{GSCT, KAIST, Korea}

\icmlcorrespondingauthor{Juhan Nam}{juhannam@kaist.ac.kr}

\icmlkeywords{Artist Recognition, Deep Convolutional Neural Network, i-vector}

\vskip 0.3in
]

\printAffiliationsAndNotice{}  % leave blank if no need to mention equal contribution

\begin{abstract}
Artist recognition is a task of modeling the artist's musical style. This problem is challenging because there is no clear standard. We propose a hybrid method of the generative model i-vector and the discriminative model deep convolutional neural network. We show that this approach achieves state-of-the-art performance by complementing each other. In addition, we briefly explain the advantages and disadvantages of each approach.

% dhk: We focus on the representation techniques for artist music style recognition which plays an important role in Music Information Retrieval system. We propose new hybrid method combining DCNN and i-vector properly by taking complementary effect. We show that this approach achieves the state-of-the-art performance on artist recognition task and briefly analyze the reason for the improvement.
\end{abstract}
%\vspace{-3mm}

\section{Introduction}
The musical style of artist is an important feature in several music information retrieval tasks such as recommending similar artists. However, since there is no clear standard for this, early approaches proposed to extract and combine various hand-crafted audio features such as timbre, harmonic contents, etc. \cite{bergstra+casagrande+erhan+eck+kegl:2006, Ellis2007Classifying}. Recent approaches focused on modeling artist by leveraging the i-vector speaker and artist recognition systems \cite{Eghbalzadeh2015IVectorsFT}.
In this paper, we adopt deep audio feature extracted from deep convolutional neural network (DCNN) combining with i-vector to alleviate the limitation of compact frame-level representation by capturing higher-level artist feature. 
%eghbalcosine}.
%\vspace{-2mm}

%\section{Methods}

\textbf{i-vector}: The i-vector is the state-of-the-art algorithm in a speaker verification \cite{Dehak2011} and also showed good performance on artist classification task \cite{Eghbalzadeh2015IVectorsFT}. We implemented i-vector using 20-dim Mel-Frequency Cepstrum Coefficients with Gaussian mixture model of size 256. We use probabilistic linear discriminant analysis (PLDA) to compute the i-vector score \cite{kenny2010bayesian}.

\textbf{DCNN}: Representation learning has been actively explored in recent years as an alternative to feature engineering \cite{Bengio:2013:RLR:2498740.2498889}. We construct the DCNN with five convolutional layer and one fully-connected layer to classify the artists using 3-second mel-spectrogram with 128 bins as input. We use the DCNN as a feature extractor and the last hidden layer (256-dim vector) as a deep audio feature. PLDA is also applied as a scoring method.

%\textbf{Hybrid}: 
The results show that i-vector and DCNN capture the characteristics of each artist differently. We also found that the two methods above are complementary to each other by showing that a hybrid approach performs better.

\vspace{-2mm}
\section{Experimental Setup}
% We validate the effectiveness of the proposed method
We conducted artist recognition on Million Song Dataset (MSD) \cite{bertin2011million} by artist verification and artist identification. 
We filtered out 20 songs for each artist which are randomly selected including various albums to prevent recording environment effects. Apart from the training data, we use 500 unseen artists for the evaluation. For evaluation, 15 songs are used to enroll each artist model and remaining 5 songs are used for testing. We aggregate the 15 track vectors to make artist model by averaging.

\textbf{Artist verification}: We compute the distance between the claimed artist model and the test feature vector. 
%The decision for verification is made by comparing the distance to a threshold. 
We evaluate the verification task in terms of equal error rate (EER), where both acceptance and rejection error rates are equal. 

\textbf{Artist identification}: There are 500 artist models and the task is choosing one of them by computing distance between the test feature vector and all artist models. We evaluate the identification task in terms of classification accuracy, which is calculated by dividing the number of correct results by the total number of test cases.

\vspace{-2mm}
\section{Results}

\subsection{The Number of Training Artists}
%\dhk{We first check how the results change as the number of the artist increases. As shown in Figure \ref{fig:reco}, both verification and identification show that the DCNN continues to improve the performance, while the i-vector converges early (with lower performance). This can be explained by the difference in data sensitivity between supervised and unsupervised learning. Since we are evaluating through classifying unseen 500 artists, DCNN, which is supervised learning, shows better performance as data increases, but i-vector, unsupervised learning, converges at around 500.}

We used increasing number of artists equally in training i-vector and DCNN to investigate how the number of artists affects the performance. Figure \ref{fig:reco} shows the experimental results of verification and identification, respectively. In both cases, the performances of DCNN are continuously improved as the training artists increase, while i-vector converges. This might be related to our experimental setting where 500 artist identity models are used in evaluation. That is, in order to discriminate a large number of artists, the supervised feature learning with DCNN also requires an equivalent or larger number of artists, accordingly. On the other hand, i-vector, which is based on unsupervised learning, is less sensitive to the number of training artists. 

\vspace{-1mm}
\subsection{Confusion Matrix}
\begin{figure}[t]
% \centering
%\hspace{-1cm}
\includegraphics[height=3.2cm,keepaspectratio]{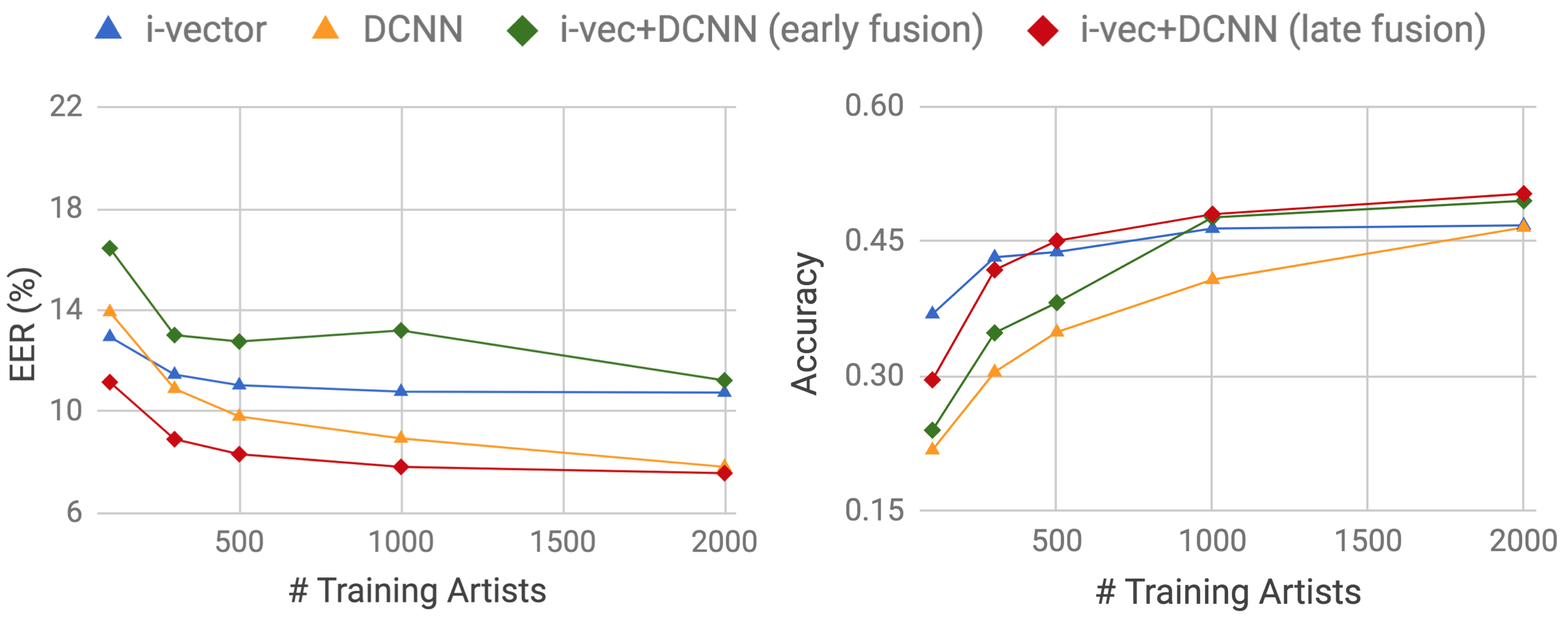} 
\begin{minipage}[t]{.5\linewidth}
\centering
\vspace{-2mm}
\subcaption{Verification}\label{veri}
\end{minipage}%
\begin{minipage}[t]{.5\linewidth}
\centering
\vspace{-2mm}
\subcaption{Identification}\label{iden}
\end{minipage}
\vspace{-5mm}
\caption[Artist Recognition]{Results of the artist recognition tasks. \label{fig:reco}}
\end{figure}

\begin{figure}[t]
 \centerline{{
 \includegraphics[width=7.5cm,height=4.3cm]{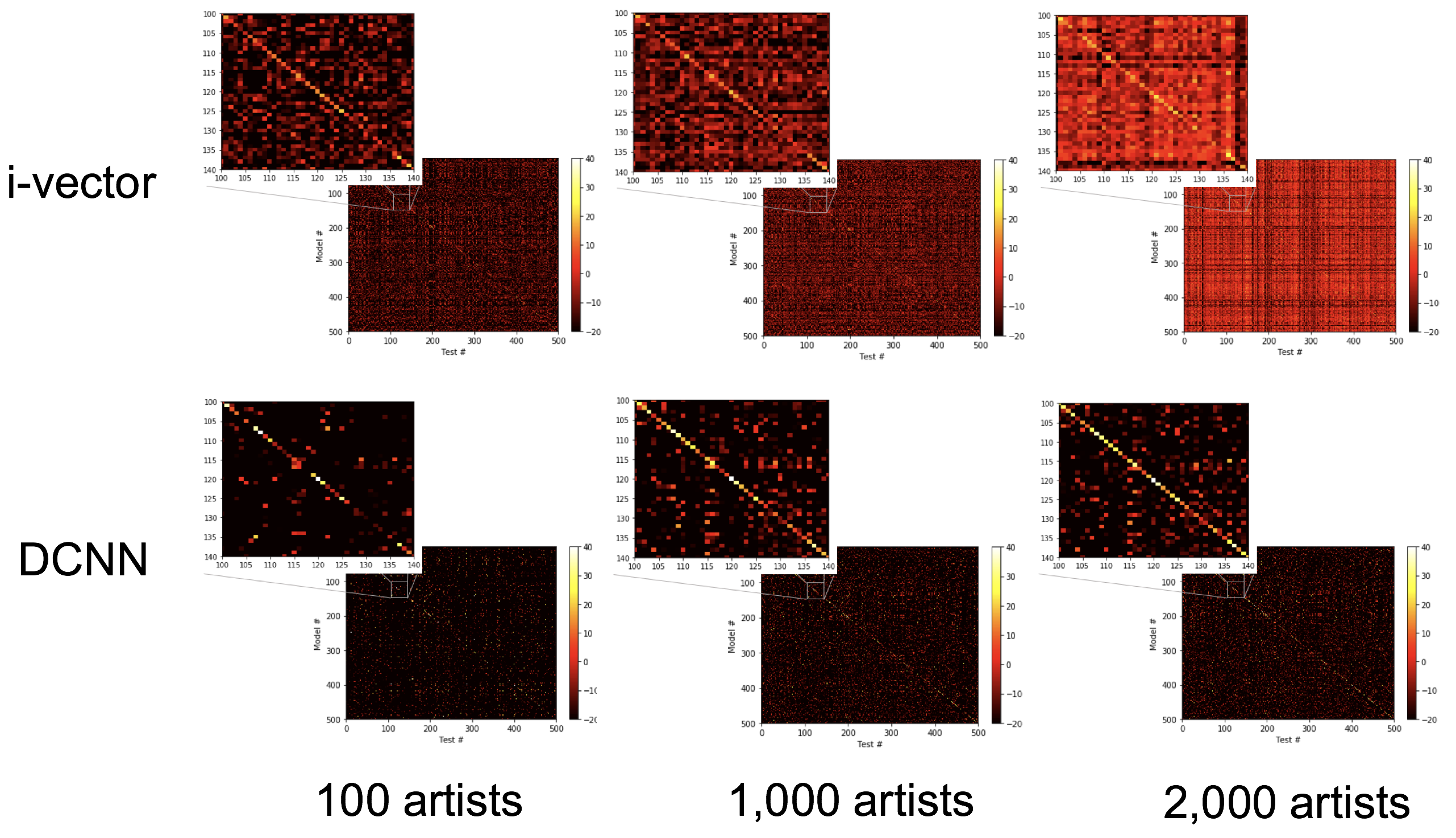}}}
 \vspace{-2mm}
 \caption{Comparison of the score matrices of i-vector and DCNN with different number of training artists.}
\label{fig:number}
\vspace{-3mm}
\end{figure}

These characteristics are also found in Figure \ref{fig:number}, which shows the score matrices with increasing the number of training artists. Each element means the similarity $S(x_i, x_j)$ where $x_i$ denotes the feature vector of $i$th artist. In this figure, we can see that, as the training artists increase, the empty portion in the middle of the diagonal line in DCNN is gradually filled. However, still, some artists' identity models are not formed well. On the other hand, i-vector can form each artist model even though the number of training artists is small. However, as the number of the training artists increases, the similarity with other artists as well as their own models increases, which makes the performance of i-vector converges. These characteristics can explain the reason why i-vector outperforms DCNN in identification task and when the number of training artists is small, whereas DCNN outperforms in verification task in Figure \ref{fig:reco}.

% \vspace{2mm}

\vspace{-1mm}
\subsection{Singer Recognition}
Because singing voice is one of the main concerns of musical pieces, and most people use the singing voice as the primary cue for recognizing a song, we also perform singer recognition to distinguish music more focusing on singing voice. We selected \textit{singers} using a CNN-based singing voice detector \cite{schluter2015exploring} by regarding the artist who has more than 20 audio clips with 70\% vocal confidence as a singer. Table \ref{tab:singer} shows the comparison results between artist and singer recognition. Compared to the artist recognition and DCNN, i-vector results are greatly improved in singer recognition. This indicates that i-vector distinguishes the human voice more clearly than music audio, and it may be related that i-vector was designed for speaker recognition.

% \begin{table}[t]
% \centering
% \resizebox{\columnwidth}{!}{\begin{tabular}{|c|c|c|c|c|c|c|}
% \hline
% \multirow{2}{*}{} & \multicolumn{3}{c|}{Verification (EER)} & \multicolumn{3}{c|}{Identification (Accuracy)} \\ \cline{2-7} 
%                   & i-vector  & DCNN  & Late fusion  & i-vector    & DCNN    & Late fusion    \\ \hline
% Artist            & 10.7850    & 8.9380       & 7.8125       & 0.4644      & 0.4076          & 0.4804         \\ \hline
% Singer            & 8.2571    & 7.6111       & 3.2414      & 0.5604      & 0.4352          & 0.7600         \\ \hline
% \end{tabular}}
% \caption{The comparison of the results between artist and singer. 1,000 artists and singers are used for training in this experiment, respectively.}
% \label{tab:singer}
% \end{table}

% \begin{table}[t]
% \centering
% \resizebox{\columnwidth}{!}{\begin{tabular}{|c|c|c|c|c|c|c|c|c|}
% \hline
% \multirow{2}{*}{} & \multicolumn{4}{c|}{Verification (EER)}   & \multicolumn{4}{c|}{Identification (Accuracy)}   \\ \cline{2-9} 
%                   & i-vec & DCNN & Early & Late & i-vec & DCNN & Early & Late \\ \hline
% Artist     &        10.7850    & 8.9380    & 13.1996    & 7.8125       & 0.4644      & 0.4076     & 0.4768    & 0.4804     \\ \hline
% Singer           & 8.2571    & 7.6111    & 10.1789   & 3.2414      & 0.5604      & 0.4352      & 0.4296   & 0.7600    \\ \hline
% \end{tabular}}
% \caption{The comparison of the results between artist and singer. 1,000 artists and singers are used for training in this experiment, respectively.}
% \label{tab:singer}
% \end{table}

\begin{table}[t]
\centering
\resizebox{\columnwidth}{!}{\begin{tabular}{|c|c|c|c|c|c|c|c|c|}
\hline
\multirow{2}{*}{} & \multicolumn{4}{c|}{Verification (EER)}   & \multicolumn{4}{c|}{Identification (Accuracy)}   \\ \cline{2-9} 
                  & i-vec & DCNN & Early & Late & i-vec & DCNN & Early & Late \\ \hline
Artist     &        10.785    & 8.938    & 13.200    & 7.813       & 0.464      & 0.408     & 0.477    & 0.480     \\ \hline
Singer           & 8.257    & 7.611    & 10.179   & 3.241      & 0.560      & 0.435      & 0.430   & 0.760    \\ \hline
\end{tabular}}
\caption{The comparison of the results between artist and singer. 1,000 artists and singers are used for training in this experiment, respectively.}
\label{tab:singer}
\end{table}
% \vspace{-2mm}

\subsection{Hybrid Methods}
\begin{figure}[t!]
 \centerline{{
 \includegraphics[width=8.4cm,height=1.68cm]{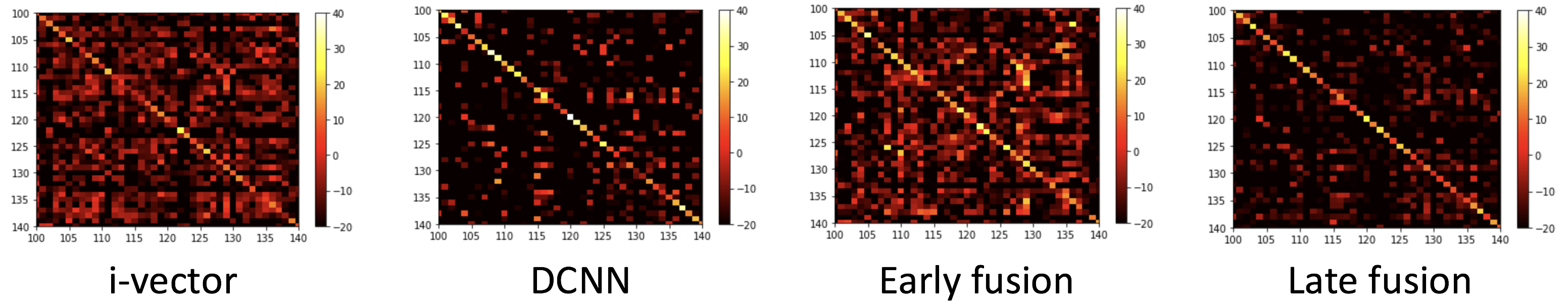}}}
\vspace{-2mm}
 \caption{Comparison of the score matrices of i-vector, DCNN, early fusion and late fusion when the number of training singers is 1,000.}
\label{fig:fusion}
\end{figure}

We also compare two hybrid methods of combining DCNN and i-vector. One is early fusion that concatenates deep audio feature and i-vector into a single feature vector before scoring, and the other is late fusion that uses the average evaluation score from both features. In Figure \ref{fig:reco}, late fusion achieves best results for all cases, whereas early fusion is generally worse than either i-vector or DCNN. In addition, the late fusion results are significantly improved in singer recognition in Table \ref{tab:singer}. From Figure \ref{fig:fusion}, we can explain the reason as early fusion seems to suppress the feature of each model by causing confusion to distinguish, while late fusion seems to take the advantages of each model and offset the disadvantages by complementing each other. A similar result can be found in audio scene classification \cite{eghbal2016cp}.

% As shown in Table {tab:singer}, late fusion achieves the best performance while early fusion gets worse than others. We can guess in figure \ref{fig:fusion} that early fusion seems to suppress the features of each method, while late fusion seems to encourage each feature well. A similar result can be found in audio scene classification \cite{eghbal2016cp}.

\section{Conclusions}
In this paper, we conducted artist recognition by verification and identification. From the results, we showed that the late fusion of deep audio feature and i-vector achieves best performance by complementing each other. We also explained the advantages and disadvantages of each approach. For future work, we will develop the aggregating method and apply the proposed method to recommend similar artists.
%delve into more clear interpretation by analyzing the captured features and we will use the hybrid method for recommending similar artists.

% In this paper, through our experiments, we showed that late fusion hybrid model is more effective in artist recognition task and the results also shows better performance than previous state-of-the-art model. We also briefly describe the contemporary effect about these hybrid model. As for future work, we will delve deeper into features and fusions for a more clear interpretation and look into the relationship with other interpretable hand-crafted features.

% \section*{Acknowledgements}


\bibliography{example_paper}

\begin{thebibliography}{9}
\providecommand{\natexlab}[1]{#1}
\providecommand{\url}[1]{\texttt{#1}}
\expandafter\ifx\csname urlstyle\endcsname\relax
  \providecommand{\doi}[1]{doi: #1}\else
  \providecommand{\doi}{doi: \begingroup \urlstyle{rm}\Url}\fi

\bibitem[Bengio et~al.(2013)Bengio, Courville, and
  Vincent]{Bengio:2013:RLR:2498740.2498889}
Bengio, Yoshua, Courville, Aaron, and Vincent, Pascal.
\newblock Representation learning: A review and new perspectives.
\newblock \emph{IEEE Trans. Pattern Anal. Mach. Intell.}, 35\penalty0 (8),
  August 2013.

\bibitem[Bergstra et~al.(2006)Bergstra, Casagrande, Erhan, Eck, and
  K{\'{e}}gl]{bergstra+casagrande+erhan+eck+kegl:2006}
Bergstra, James, Casagrande, Norman, Erhan, Dumitru, Eck, Douglas, and
  K{\'{e}}gl, Bal{\'{a}}zs.
\newblock Aggregate features and adaboost for music classification.
\newblock \emph{Machine Learning}, 65:\penalty0 473--484, December 2006.

\bibitem[Bertin-Mahieux et~al.(2011)Bertin-Mahieux, Ellis, Whitman, and
  Lamere]{bertin2011million}
Bertin-Mahieux, Thierry, Ellis, Daniel~PW, Whitman, Brian, and Lamere, Paul.
\newblock The million song dataset.
\newblock In \emph{Ismir}, volume~2, pp.\ ~10, 2011.

\bibitem[Dehak et~al.(2011)Dehak, Kenny, Dehak, Dumouchel, and
  Ouellet]{Dehak2011}
Dehak, N., Kenny, P.~J., Dehak, R., Dumouchel, P., and Ouellet, P.
\newblock Front-end factor analysis for speaker verification.
\newblock \emph{IEEE Transactions on Audio, Speech, and Language Processing},
  pp.\  788--798, 2011.

\bibitem[Eghbal-Zadeh et~al.(2015)Eghbal-Zadeh, Lehner, Schedl, and
  Widmer]{Eghbalzadeh2015IVectorsFT}
Eghbal-Zadeh, Hamid, Lehner, Bernhard, Schedl, Markus, and Widmer, Gerhard.
\newblock I-vectors for timbre-based music similarity and music artist
  classification.
\newblock In \emph{Proceedings of the International Society for Music
  Information Retrieval Conference (ISMIR)}, pp.\  554--560, 2015.

\bibitem[Eghbal-Zadeh et~al.(2016)Eghbal-Zadeh, Lehner, Dorfer, and
  Widmer]{eghbal2016cp}
Eghbal-Zadeh, Hamid, Lehner, Bernhard, Dorfer, Matthias, and Widmer, Gerhard.
\newblock {CP-JKU} submissions for dcase-2016: A hybrid approach using binaural
  i-vectors and deep convolutional neural networks.
\newblock \emph{IEEE AASP Challenge on Detection and Classification of Acoustic
  Scenes and Events (DCASE)}, 2016.

\bibitem[Ellis(2007)]{Ellis2007Classifying}
Ellis, Daniel~PW.
\newblock Classifying music audio with timbral and chroma features.
\newblock In \emph{ISMIR 2007: Proceedings of the 8th International Conference
  on Music Information Retrieval: September 23-27, 2007, Vienna, Austria}, pp.\
   339--340. Austrian Computer Society, 2007.

\bibitem[Kenny(2010)]{kenny2010bayesian}
Kenny, Patrick.
\newblock Bayesian speaker verification with heavy-tailed priors.
\newblock In \emph{Odyssey}, pp.\ ~14, 2010.

\bibitem[Schl{\"u}ter \& Grill(2015)Schl{\"u}ter and
  Grill]{schluter2015exploring}
Schl{\"u}ter, Jan and Grill, Thomas.
\newblock Exploring data augmentation for improved singing voice detection with
  neural networks.
\newblock In \emph{Proceedings of the International Society for Music
  Information Retrieval Conference (ISMIR)}, pp.\  121--126, 2015.

\end{thebibliography}
\bibliographystyle{icml2018}

\end{document}